\documentclass[preprint,prc,aps,showpacs,groupedaddress,superscriptaddress,floatfix]{revtex4}
\usepackage{amsmath}
\usepackage{amssymb}

\begin{document}

\title{\textbf{Non-relativistic quark-antiquark potential: spectroscopy of
heavy-quarkonia and exotic SUSY quarkonia}}
\author{Sameer M. Ikhdair}
\email[E-mail: ]{sameer@neu.edu.tr; sikhdair@neu.edu.tr}
\affiliation{Department of Physics, Near East University, Nicosia, Cyprus, Turkey}
\author{Ramazan Sever}
\email[E-mail: ]{sever@metu.edu.tr}
\affiliation{Department of Physics, Middle East Technical University, 06531, Ankara,Turkey}
\date{\today}

\begin{abstract}
The experiments at LHC have shown that the SUSY (exotic) bound states are
likely to form bound states in an entirely similar fashion as ordinary
quarks form bound states, i.e., quarkonium. Also, the interaction between
two squarks is due to gluon exchange which is found to be very similar to
that interaction between two ordinary quarks. This motivates us to solve the
Schr\"{o}dinger equation with a strictly phenomenological static
quark-antiquark potential: $V(r)=-Ar^{-1}+\kappa \sqrt{r}+V_{0}~$\ using the
shifted large $N$-expansion method to calculate the low-lying spectrum of a
heavy quark with anti-sbottom\textbf{\ }$(c\overline{\widetilde{b}},b%
\overline{\widetilde{b}})$ and sbottom with anti-sbottom $(\widetilde{b}%
\overline{\widetilde{b}})$ bound states with $m_{\widetilde{b}}$ is set
free. To have a full knowledge on spectrum, we also give the result for a
heavier as well as for lighter sbottom masses. As a test for the reliability
of these calculations, we fix the parameters of this potential by fitting
the spin-triplet $(n^{3}S_{1})$ and center-of-gravity $l\neq 0$ experimental
spectrum of the ordinary heavy quarkonia $c\overline{c},c\overline{b}$ and $b%
\overline{b}$ to few $\mathrm{MeV.}$ Our results are compared with other
models to gauge the reliability of these predictions and point out
differences.

\end{abstract}

\maketitle

\bigskip

\section{Introduction}

\noindent Weak-scale Supersymmetric Standard Model (SSM) is the leading
candidate for physics beyond the Standard Model (SM) [1,2]. Supersymmetry
(SUSY) is built on a solid theoretical and mathematical foundation. It is
also well-motivated as an elegant solution to the gauge hierarchy problem
and has merits of gauge coupling unification, dynanical electroweak symmetry
and providing a legitimate candidate for dark matter. SUSY predicts the
existence of a super partner called SUSY particles (sparticles)
corresponding to each ordinary particle of SM. These sparticles should be
accessible at the exist and constructing colliders such as $\mathrm{Tevatron}
$ and $\mathrm{LHC}$. Over the past years, great effort has been made to
search for such sparticles. So far, no direct signal for SUSY has been
observed and some lower mass bounds have been established for sparticles.
The experimental results at $\mathrm{LEP}$ [3-8,9,10] and $\mathrm{Tevatron}$
[11-14], squarks must be heavier than about $\mathrm{100}$ $\mathrm{GeV.}$
However, most experimental searches for sparticles are performed with
model-dependent assumptions and rely on a large missing energy cut. A
long-lived light SUSY bottom quark (sbottom), $\widetilde{b},$ and its
anti-sbottom, $\overline{\widetilde{b}},$ with a mass $m_{\widetilde{b}}$
close to $m_{b}$ ($\mathrm{\thicksim 4.9}$ $\mathrm{GeV}$), roughly half the
$\Upsilon (1S)$ mass, has not been excluded by experiments [15,16]. Hence, a
light sbottom and its anti-sbottom, are not excluded so far partly because
of the $\mathrm{ALEPH}$ collaboration indication [3-8] and partly because of
interesting scenario to explain the excess of $b\overline{b}$ pair
production in hadron collisions than theoretical prediction by a factor two.
Some analyses [15] showed that if the light $\widetilde{b}$ is an
appropriate admixture of left-handed and right-handed sbottom quark, its
coupling to $Z$ boson can be small enough to avoid $\mathrm{LEP}$-$\mathrm{I}
$ $\mathrm{Z}$ decay bounds. In addition, a scenario with light gluino and
long-lived light sbottom with mass close to the bottom quark was proposed in
[16] with which the excess of measured $b\widetilde{b}$ pair production in
hadron collision over QCD theoretical prediction by a factor two is
explained successfully (cf. [17-19]). The data about $b\overline{b}$ pair
production in hadron collision given by CDF and D0 can be explained well by
QCD theoretical production: e.g., we can learn the details from the web
address in [17-19]. The $\mathrm{CLEO}$ exclusion of a $\widetilde{b}$ with
mass $\mathrm{3.5}$ to $\mathrm{4.5}$ $\mathrm{GeV}$ [20] can also be loosed
even avoided, since their analysis depends on the assumption for
semi-leptonic decays of the light sbottom. Moreover, since sbottom is a
scalar, based on the spin freedom counting only, its pair production rate at
collision will be smaller than the bottom quark by a factor four, so the
sbottom samples must be rarer than those of bottom quark in experiments.

In contrary, it is interesting to point out that some experiments seemingly
favor such a light sbottom. The $\mathrm{ALEPH}$ collaboration has reported
experimental hints for a light sbottom with a mass around $\mathrm{4}$ $%
\mathrm{GeV}$ and lifetime of $\mathrm{1}$ $\mathrm{ps}$ [21]. A recent
analysis of old anomaly in the $\mathrm{MARK}$-$\mathrm{I}$ data for cross
section of $\mathrm{e}^{+}\mathrm{e}^{-}\mathrm{\rightarrow }$hadrons shows
that the existence of such a light sbottom can bring the measured cross
section into agreement with the theoretical prediction [22]. As mentioned by
Berger\textit{\ et al}. [16], a light-gluino analysis was done by Baer
\textit{et al}. [23] in which the gluino is assumed $\mathrm{LSP}$. Cheung
and Keung [24] modified the analysis [23] by letting the light gluino decay
into $b$ and $\widetilde{b}$ and study the possible constraint and
implication at $\mathrm{LSP}$\textrm{.} Therefore, the light gluino and
light sbottom scenario will certainly give rise to other interesting
signals, e.g., decay of $\chi _{b}$ into the light sbottom [25], enhancement
of $t\overline{t}b\overline{b}$ production at hadron colliders [26], decay
of $\Upsilon $ into a pair of light sbottoms [27] and flavor-changing
effects in radiative decays of $B$ mesons [28].

The phenomenology of a very light sbottom has been studied by many authors
recently [24-26,28-32]. If such a light sbottom indeed exist, new meson-like
bound states is formed by a pair of the sbottom and anti-sbottom ($%
\widetilde{b}\overline{\widetilde{b}}$) and fermion-like ones by an ordinary
quark with anti-sbottom $(q\overline{\widetilde{b}}$) $(e.g.,$ heavy quark $%
q=$ $c,b)$ may be also formed.

Some quarkonium binding systems like $c\overline{c},$ $b\overline{b}$ and $c%
\overline{b}$ have been studied with encouraging success [33-35], in the
framework of the potential model using a strictly phenomenological static
heavy quark-antiquark potential belonging to the generality $%
V(r)=-Ar^{-\alpha }+\kappa r^{\beta }+V_{0},$ $(\alpha =1,\beta =1/2).$
Hence, the parameters of this potential are fixed by fitting the
experimentally measured triplet $\mathrm{S}$-states and the
center-of-gravity (c.o.g.) non $\mathrm{S}$-states of $c\overline{c}$ and $b%
\overline{b}$ spectra to their calculated levels and taking into
consideration their hyperfine splittings in the framework of
non-relativistic quarkonium model. It has been found that the potential
description is flavor-independent, i.e., the same potential describes
equally well the $c\overline{c}$ and the $b\overline{b}$ systems. Therefore,
if we take the potential parameters obtained from the fitting of the mass
spectra $c\overline{c}$ and the $b\overline{b}$ systems, we may predict the
Schr\"{o}dinger bound-state masses of the exotic states. We take the very
same values of potential parameters used for the observed bound states of
ordinary quarks to predict the unknown exotic squark bound states. It is
well-known that the interaction between two squarks is due to the gluon
exchange [36] which is found to be very similar to that interaction between
two ordinary quarks, and an interaction due to Higgs particle exchange [37].
Thus, the gluon exchange interaction in squarkonium motivates us to use very
similar parameter set of present potential model as in quarkonium [33-35].

Over the past years, the experiments at LHC have shown that the exotic bound
states are likely to form bound states in an entirely similar fashion as
ordinary quarks form bound state, i.e., quarkonium. The long-lived sbottom
is not excluded by conventional searches and an analysis should be done to
verify that there are no additional constraints on the allowed range of
sbottom masses and lifetimes. In addition, it is well-known that the
interaction between two squarks is due to gluon exchange which is found to
be very similar to that interaction between two ordinary quarks, and an
interaction due to Higgs particle exchange. Thus, the gluon exchange
interaction in squarkonium motivates us to use very similar parameter set of
present potential model as in quarkonium. The purpose is to calculate the
spectroscopy of $(Q\overline{\widetilde{b}})$ $Q=c,b$ and ($\widetilde{b}%
\overline{\widetilde{b}}$) in terms of potential model with Coulomb plus
square-root potential [33-35] in which the parameters are fixed by heavy
quarkonia ($c\overline{c}$) and ($b\overline{b}$) with sbottom mass, $m_{%
\widetilde{b}}$ is set as a free parameter. Furthermore, in order to have a
full knowledge on such a spectrum, we also give the result for a heavier
sbottom masses from $\mathrm{3.0}$ $\mathrm{GeV}$ to $\mathrm{60.0}$ $%
\mathrm{GeV.}$

The outline of this paper is as following: In section 2, we first review the
analytic solution of the Schr\"{o}dinger equation for non-equal mass case
for the spin-triplet $S$-states and c.o.g $\ $(center of gravity) $P$- and $%
D $-states. In section 3, we briefly present the squarkonium production
through the leptonic decay width and through the $Z^{0}$ decay. We present
the observed states and reproduce the calculated spectrum of the $c\overline{%
c},$ $c\overline{b}$ and $b\overline{b}$ spectra in section 4. Also, the
corresponding triplet $S$-states and the c.o.g. $l=1,2$ \ for an ordinary SM
heavy quark with anti-sbottom\textbf{\ }$(c\overline{\widetilde{b}},b%
\overline{\widetilde{b}})$ and sbottom with anti-sbottom $\widetilde{b}%
\overline{\widetilde{b}}$ meson-like binding system with sbottom mass, $m_{%
\widetilde{b}}$ is set as a free parameter. The conclusions are also given
in section 5. \

\section{\noindent spin-averaged binding mass spectrum}

\noindent We limit our discussion to the following generality of potentials
[33,38--49]:

\begin{equation}
V(r)=-Ar^{-\alpha }+\kappa r^{\beta }+V_{0},~\alpha ,\beta >0
\end{equation}%
where $A$ and $\kappa $ are positive constants whereas $V_{0}$ is taking any
sign. These static quarkonium potentials are monotone nondecreasing, and
concave functions satisfying the condition [41-49]%
\begin{equation}
V^{\prime }(r)>0\text{ \ and }V^{\prime \prime }(r)\leq 0.
\end{equation}%
At least ten potentials of this generality, but with various values of the
parameters, have been proposed in the literature (see, for example, Ref.
[33] and references therein). Motyka and Zalewiski [34,35] have also
explored the quality of fit in the region $0\leq \alpha \leq 1.2,$ $0\leq
\beta \leq 1.1$ of the $\alpha ,\beta $ plane reasonably well with
coordinates $\alpha =1,\beta =0.5$. Therefore, the five parameters $%
(A,\kappa ,V_{0},m_{c},m_{b})$ are fixed, in fitting the $c\overline{c}$ and
the $b\overline{b}$ experimental triplet states in flavor-independent model,
as the below values:
\begin{equation}
V(r)=-\frac{0.325250}{r}+0.70638\sqrt{r}-0.78891,
\end{equation}%
with fitted quark masses:%
\begin{equation}
m_{c}=1.3959\text{ }GeV,\text{ }m_{b}=4.8030\text{ }GeV,\text{ \ \ \ \ }
\end{equation}%
where $V(r),$ $\sqrt{r}$ and $r^{-1}$ are all in units of $\mathrm{GeV}.$
Notice that for the $c$ quark, $m_{c}$ is roughly half the $J/\psi $ mass
and for $b$ quark it is roughly half of the $\Upsilon (1S)$ mass. Thus, the
potential model (3) is convincing as it approaches to the perturbative QCD
formula in the short-distance region and approached to the confining
potential in the long-distance region. Consequently, in short-distance region%
$,$ this potential involves the $r^{-1}$ \ (Coulombic part) corresponding to
one gluon exchange which is approaching \ to the perturbative QCD formula.
The linear confinement part of the potential is $\thicksim $ $r,$ as in
Cornell potential [50,51], is not seen. Such a linearly rising potential
[50,51] is capable of confining quarks permanently and it can give rise to
spectrum of particles containing light quarks in rough accord with
experiment [52,53]. Probably the heavy quarkonia like $b\overline{b},$ $c%
\overline{b}$ $(B_{c}$ meson$)$ and $\widetilde{b}\overline{\widetilde{b}}$
are too small to reach sufficiently far into the asymptotic region of linear
confinement. On the other hand, the charmonium $c\overline{c}$ is too large
to reach sufficiently far into the confining potential in the large distance
particularly for excited states near and above the open flavor threshold.
Perhaps a more flexible potential would exhibit the linear part, but one may
be observing an effect of the expected screening of the interaction between
the heavy quarks by the light sea quarks [54].

We choose the corresponding spin triplet c.o.g. states for the practical
reasons that the masses of the spin singlet pseudoscalars for the
bottomonium $b\overline{b}$ are currently unknown or very poorly measured
[55] and the two unknown charmonium $c\overline{c}$ states in the $\mathrm{3S%
}$ and $\mathrm{4S}$ multiplets are the $\mathrm{3}^{1}\mathrm{S}_{0}$ and $%
\mathrm{4}^{1}\mathrm{S}_{0}$ pseudoscalars.

For a system of two composite particles, we shall consider the $\mathrm{D}$%
-dimensional space Schr\"{o}dinger equation for any spherically symmetric
central potential in (3). Using $\psi (\mathbf{r})=Y_{l,m}(\theta ,\phi
)u(r)/r^{(D-1)/2},$ it is straightforward to find the $l\neq 0$ radial wave
equation (in the usual units $\hbar =c=1):$

\begin{equation}
\left\{ -\frac{1}{4\mu }\frac{d^{2}}{dr^{2}}+\frac{[\overline{k}-(1-a)][%
\overline{k}-(3-a)]}{16\mu r^{2}}+V(r)\right\} u(r)=E_{n,l}u(r),\text{ }%
\overline{k}=D+2l-a,
\end{equation}%
where $u(r)$ is the radial wave function, $E_{n,l}$ is the Schr\"{o}dinger
binding energy of meson and $a$ is a proper shift. We follow the shifted $%
1/N $ or $1/\overline{k}$ expansion method by defining
\begin{equation}
V(x(r_{0}))\;=\overset{\infty }{\underset{m=0}{\sum }}\left( \frac{%
d^{m}V(r_{0})}{dr_{0}^{m}}\right) \frac{\left( r_{0}x\right) ^{m}}{m!Q}%
\overline{k}^{(4-m)/2},\text{ }Q=\overline{k}^{2},
\end{equation}%
and the binding energy expansion%
\begin{equation}
E_{n,l}\;=\overset{\infty }{\underset{m=0}{\sum }}\frac{\overline{k}^{(2-m)}%
}{Q}E_{m},
\end{equation}%
where $x=\overline{k}^{1/2}(r/r_{0}-1),$ $r_{0}$ is an arbitrary point where
the Taylor's expansions is being performed around. Following Refs.
[33,38-49], we give the necessary expressions for calculating the binding
energies to the third order:%
\begin{equation}
E_{0}=V(r_{0})+\frac{\beta }{16\mu },
\end{equation}%
\begin{equation}
E_{1}=\beta \left[ \left( n_{r}+\frac{1}{2}\right) \omega -\frac{(2-a)}{8\mu
}\right] ,
\end{equation}%
\begin{equation}
E_{2}=\beta \left[ \frac{(1-a)(3-a)}{16\mu }+\alpha ^{(1)}\right] ,
\end{equation}%
\begin{equation}
E_{3}=\beta \alpha ^{(2)},\text{ }\beta =\left( \frac{\overline{k}}{r_{0}}%
\right) ^{2},
\end{equation}%
where $\alpha ^{(1)}$ and $\alpha ^{(2)}$ are two useful expressions given
by Imbo \textit{et al. }[56-58] and also the parameter $\overline{k}$ is%
\begin{equation}
\overline{k}=\sqrt{8\mu r_{0}^{3}V^{\prime }(r_{0})}.
\end{equation}%
Hence, the total binding energy of the three-dimensional ($D=3$) Schr\"{o}%
dinger equation to the third order is
\begin{equation}
E_{n,l}=V(r_{0})+\frac{1}{2}r_{0}V^{\prime }(r_{0})+\frac{1}{r_{0}^{2}}\left[
\frac{(1-a)(3-a)}{16\mu }+\alpha ^{(1)}+\frac{\alpha ^{(2)}}{\overline{k}}%
+O\left( \frac{1}{\overline{k}^{2}}\right) \right] ,
\end{equation}%
and the shift parameter is%
\begin{equation}
a=2-(2n_{r}+1)\left[ 3+\frac{r_{0}V^{\prime \prime }(r_{0})}{V^{\prime
}(r_{0})}\right] ^{1/2}.
\end{equation}%
The root, $r_{0},$ in Eqs. (13)-(14) can be found through the relation:
\begin{equation}
1+2l+(2n_{r}+1)\left[ 3+\frac{r_{0}V^{\prime \prime }(r_{0})}{V^{\prime
}(r_{0})}\right] ^{1/2}=\left[ 8\mu r_{0}^{3}V^{\prime }(r_{0})\right]
^{1/2},\text{ }n_{r},l=0,1,2,3,\cdots ,
\end{equation}%
where the radial number $n_{r}=n-1$ with $n=1,2,3,\cdots $ is the principal
quantum number. Once $r_{0}$ is determined through Eq. (15), hence finding
the Schr\"{o}dinger binding energy of any quarkonium system from Eq. (13)
becomes relatively simple and straightforward. Finally, the corresponding
ordinary or exotic bound states become%
\begin{equation*}
M(q_{i}\overline{q}_{j})_{nl}=m_{q_{i}}+m_{q_{j}}+2E_{n,l},
\end{equation*}%
\begin{equation}
M(q_{i}\overline{\widetilde{q}}_{j})_{nl}=m_{q_{i}}+m_{\widetilde{q}%
_{j}}+2E_{n,l},
\end{equation}%
where $m_{q_{i}}$ and $m_{q_{j}}$ are the composite masses of the quark with
antiquark and $m_{\widetilde{q}_{i}}$ and $m_{\overline{\widetilde{q}}_{j}}$
squark with anti-squark Details of the model and the method of solution may
be found in Refs. [33,38-48].

Now let us turn to the investigation of the spin-spin interaction. It is
well-known that the system under study is a nonrelativistic, the treatment
is based on the Schr\"{o}dinger equation with a Hamiltonian [59,60]
\begin{equation}
H_{o}=-\frac{\triangledown ^{2}}{2\mu }+V(r)+V_{SS},
\end{equation}%
where $V_{SS}$ is the spin-spin contact hyperfine interaction which is one
of the spin-dependent terms predicted by one-gluon exchange (OGE) forces
[59,60]. Recently, the spin-spin part, in momentum space $(q=\mu ),$ was
found to be [61-63]%
\begin{equation}
V_{SS}(m_{1},m_{2},q)=\frac{\mathbf{s}_{1}.\mathbf{s}_{2}}{3m_{1}m_{2}}%
g_{s}^{2}(q)\left[ \frac{N_{c}^{2}-1}{N_{c}}%
c_{3}(q,m_{1})c_{3}(q,m_{2})-6N_{c}d(q)\right] ,
\end{equation}%
with Wilson coefficient%
\begin{equation}
c_{3}(q,m)=\left( \frac{\alpha _{s}(q)}{\alpha _{s}(m)}\right) ^{-9/25}\text{
and \ }d(q)=\frac{N_{c}^{2}-1}{8N_{c}^{2}}\left( \frac{\alpha _{s}(m_{2})}{%
\alpha _{s}(m_{1})}\right) ^{-9/25}\left[ 1-\left( \frac{\alpha _{s}(q)}{%
\alpha _{s}(m_{2})}\right) ^{-18/25}\right] ,
\end{equation}%
where $N_{c}$ is the number of colors, $g_{s}(q)$ is the running coupling
constant [61-65]. The formula (19) improves upon the one-loop perturbative
calculation in two important respects: (i) it is independent of $\mu $ and
(ii) also includes the higher order logarithmic terms.

If the coefficients are calculated at tree level; i.e., $c_{3}(\mu ,m)=1,$ $%
d(\mu )=0,$ the potential reduces to the Eichten-Feinberg result [66,67].
And if these coefficients are expanded to order $\alpha _{s}(\mu )$ then
reduced to a one-loop quarkonium spin-spin interaction in the
nonrelativistic case [68-70] which is responsible for the hyperfine
splitting of the mass levels [71-84]%
\begin{equation}
V_{SS}\longrightarrow V_{\mathrm{HF}}=\frac{32\pi \alpha _{s}}{9m_{q_{i}}m_{%
\overline{q}_{j}}}\delta ^{3}(\mathbf{r})\left( \mathbf{s}_{1}.\mathbf{s}%
_{2}-\frac{1}{4}\right) ,
\end{equation}%
adapted from the Breit-Fermi Hamiltonian. The number $\frac{1}{4}$
substituted from the product of the spins corresponds to the recent
assumption that the unperturbed nonrelativistic Hamiltonian gives the energy
of the triplet states. Since for the states with orbital angular momentum $%
L>0$ the wave function vanishes at the origin, the shift affects only the $%
\mathrm{S}$ states. Thus, the only first order effect of this perturbation
is to shift to the pseudoscalar $^{1}S_{0}$ states down in energy:%
\begin{equation}
\Delta E_{\mathrm{HF}}=\frac{32\pi \alpha _{s}}{9m_{q_{i}}m_{\overline{q}%
_{j}}}\left\vert \psi (0)\right\vert ^{2},
\end{equation}%
with the wave function at the origin is calculated by using the expectation
value of the potential derivative via [33,41-46,49,82-84]%
\begin{equation}
\left\vert \psi (0)\right\vert ^{2}=\frac{\mu }{2\pi }\left\langle \frac{%
dV(r)}{dr}\right\rangle .
\end{equation}%
An application of the last formula needs the value of the wavefunction at
the origin. This can be achieved by solving the Schr\"{o}dinger equation
with the nonrelativistic Hamiltonian and the coupling constant. In such an
approach, the QCD strong coupling constant $\alpha _{s}(4\mu ^{2}),$ on the
renormalization point $\mu ^{2}$ is not an independent parameter. It can be
connected (in the $\overline{MS}$ renormalization scheme) through the
two-loop relation [34,35,85-87]
\begin{equation}
\alpha _{s}(\mu ^{2})=\frac{2\pi }{\beta _{0}}\frac{\eta ^{2}-1}{\ln (\eta )}%
,\text{ }\eta =\frac{2\mu }{\Lambda _{\overline{MS}}^{(n_{f})}},
\end{equation}%
where $\beta _{0}=11-\frac{2}{3}n_{f}.$ Like most other authors (see, for
example, Refs. [33-35,72-75]), the strong coupling constant $\alpha
_{s}(m_{c}^{2}),$ is fitted to the experimental charmonium hyperfine
splitting numbers $\Delta _{\mathrm{HF}}(\mathrm{1S,}$exp$)\thickapprox
\mathrm{116.5\pm 1.2}$ $\mathrm{MeV}$ and $\Delta _{\mathrm{HF}}(\mathrm{2S,}
$exp$)\thickapprox \mathrm{48.1\pm 4}$ $\mathrm{MeV}$ [41-46,55], which
yields%
\begin{equation}
\alpha _{s}(m_{c}^{2})=0.254.
\end{equation}%
Knowing the coupling at the scale $m_{c}^{2},$ we obtain the couplings at
other scales as follows. The number of flavours $(n_{f})$ is put equal to
three for $4\mu ^{2}\leq m_{c}^{2}$ (we are not interested in the region $%
4\mu ^{2}\leq m_{s}^{2}),$ equal to four for $m_{b}^{2}\geq 4\mu ^{2}\geq $ $%
m_{c}^{2}$ and equal to five for $4\mu ^{2}\geq $ $m_{b}^{2}$ (we are not
interested in the region $4\mu ^{2}\geq $ $m_{t}^{2}).$ Then the value of\ $%
\alpha _{s}(4\mu ^{2}=m_{c}^{2})$ from (23) is used to calculate $\Lambda _{%
\overline{MS}}^{(n_{f}=3)}$ and $\Lambda _{\overline{MS}}^{(n_{f}=4)}.$
Using the known value of $\Lambda _{\overline{MS}}^{(n_{f}=4)}$ and Eq. (23)
we find the value%
\begin{equation}
\alpha _{s}(m_{b}^{2})=0.200\text{ and }\alpha _{s}(4\mu _{cb}^{2})=0.224,
\end{equation}%
and%
\begin{equation}
\alpha _{s}(m_{\widetilde{b}}^{2})=\alpha _{s}(m_{b}^{2})\text{ and }\alpha
_{s}(4\mu _{c\widetilde{\overline{b}}}^{2})=\alpha _{s}(4\mu _{cb}^{2}).
\end{equation}%
We follow these analysis in order to calculate the experimental binding
masses of the heavy quarkonia $c\overline{c},$ $c\overline{b}$ and $b%
\overline{b}.$ At the end, it is worth to note that the sbottom $\widetilde{b%
}$ is a scalar, there is no spin-spin interaction (hyperfine splitting) for (%
$Q\overline{\widetilde{b}}$) and ($\widetilde{b}\overline{\widetilde{b}}$),
where $Q=c,b.$

\section{Squarkonium Production}

\subsection{Production Through the Leptonic Decay}

The squarks might be discovered by detecting ordinary quark-antiquark bound
states as resonances at LHC. (This is of course one of the main ways of
studying charm and bottom quarks). The bound states of squarks are narrow
resonances depend primarily on their leptonic widths (unless the squark
itself has a very large width). Therefore, our aim will be to compute $%
\Gamma _{e}$. The leptonic decay widths of the heavy quarkonia and
squarkonium are proportional to the squares of the wave functions at the
origin. Therefore, they are significant only for the $S$ states. To compute
the decay rate for this process, we use non-relativistic bound state
techniques since the squarks are expected to be very heavy objects (massive
particles) and therefore their bound states are not relativistic systems.
For the $c\overline{c}$ and $b\overline{b}$ quarkonium and $\widetilde{b}%
\overline{\widetilde{b}}$ sbottomonium systems, we shall consider the decays
of the $n^{3}S_{1}$ (vector, $J^{PC}=1^{--}$) states decay into charged
lepton pairs, e.g. $e^{+}e^{-}$ pairs are usually calculated from the QCD
corrected Van Royen-Weiskopf formula%
\begin{equation}
\Gamma _{e}(n^{3}S_{1}\rightarrow \overline{l}l)=16\pi \alpha ^{2}e_{q}^{2}%
\frac{\left\vert \psi (0)\right\vert ^{2}}{M_{V}^{2}}\left( 1-\frac{16\alpha
_{s}(m_{q}^{2})}{3\pi }\right) ,
\end{equation}%
where $\left\vert \psi (0)\right\vert $ is the bound state radial
wavefunction at the origin, $M_{V}$ the mass of the bound triplet (vector)
state, $\alpha $ the fine-structure constant and $e_{q}$ the charge of the
quark in units of the electron charge. In the computation we have taken for
ordinary quarks $e_{c}=\frac{2}{3}$ and $e_{b}=-\frac{1}{3}$, however, for
squarks $e_{\widetilde{b}}=\frac{1}{3}.$ For the $c\overline{b}$ quarkonium
and $c\overline{\widetilde{b}},b\overline{\widetilde{b}}$ squarkonium, we
consider the decays of the $n^{1}S_{0}$ (pseudoscalar) states into $\tau \nu
_{\tau }$ pairs. Since the probability of such decays contains as a factor
the square of the lepton mass, the decays into lighter leptons are much less
probable [33-35]. For vector mesons containing light quarks (squarks) this
formula leads to paradoxes (cf., e.g., Ref. [33] and references therein).
For quarkonia, however, the main problem seems to be the QCD correction.
Thus, in order to get quantitative predictions it is necessary to include
higher order corrections which are not known. In order to estimate the
missing terms we tried two simple forms. Exponentialization of the first
correction%
\begin{equation}
C_{1}(\alpha _{s}(m_{q}^{2}))=\exp \left( -\frac{16\alpha _{s}(m_{q}^{2})}{%
3\pi }\right) ,
\end{equation}%
and Padeization%
\begin{equation}
C_{2}(\alpha _{s}(m_{q}^{2}))=\frac{1}{1+\frac{16\alpha _{s}(m_{q}^{2})}{%
3\pi }}.
\end{equation}%
We use the average of these two estimates as our estimate of the QCD
correction factor extended to higher orders. The difference between $C_{1}$
and $C_{2}$ is our crude evaluation of the uncertainty of this estimate.
Further, we have the relation%
\begin{equation}
\Gamma _{e}(n^{3}S_{1}\rightarrow \overline{l}l)=\frac{9}{8}\frac{4m_{q}^{2}%
}{M_{V}^{2}}\frac{\alpha ^{2}e_{q}^{2}}{\alpha _{s}(m_{q}^{2})}C_{av}\Delta
E_{\mathrm{HF}},
\end{equation}%
where $C_{av}$ is the averaged QCD correction factor. With our choice of
parameters this formula reduces to%
\begin{equation}
\Gamma _{e}(n^{3}S_{1}\rightarrow \overline{l}l)=F(q)\frac{4m_{\widetilde{q}%
}^{2}}{M_{V}^{2}}\Delta E_{\mathrm{HF}},
\end{equation}%
where $m_{q}$ ($m_{\widetilde{q}}$) is the quark (squark) mass and $%
F(c)=7.07\times 10^{-5}$ and $F(b)=F(\widetilde{b})=2.43\times 10^{-5},$ see
Eq. (34). The leptonic width is a small fraction of the total width.
Experimentally, for narrow resonances (whose width is much smaller than the
beam-energy spread) one measures the integrated area of the resonance
cross-section. For a Breit-Wigner-type resonance this is connected to the
leptonic width by the formula [88]%
\begin{equation}
\int \sigma _{res}dE=\frac{2\pi ^{2}(2J+1)}{M_{V}^{2}}\frac{\Gamma
_{e}\Gamma _{h}}{\Gamma },
\end{equation}%
and $\Gamma _{h}/\Gamma \simeq 1$ if the hadronic width predominates. The
formula for the leptonic widths of the pseudoscalar $c\overline{b}$
quarkonium reads%
\begin{equation}
\Gamma _{\tau \nu _{\tau }}=\frac{G^{2}}{8\pi }f_{B_{c}}^{2}\left\vert
V_{cb}\right\vert ^{2}M_{B_{c}}m_{\tau }^{2}\left( 1-\frac{m_{\tau }^{2}}{%
M_{B_{c}}^{2}}\right) ^{2},
\end{equation}%
where $G$ is the Fermi constant, $V_{cb}\approx 0.04$ is the element of the
Cabibbo-Kobayashi-Masakawa matrix and the decay constant $f_{B_{c}}$ is
given by the formula (cf., e.g., Ref. [33] and references therein)%
\begin{equation}
f_{B_{c}}^{2}=\frac{12\left\vert \psi (0)\right\vert ^{2}}{M_{B_{c}}^{2}}%
\overline{C}^{2}(\alpha _{s}),
\end{equation}%
where $\overline{C}(\alpha _{s})$ is QCD correction factor. Formally this
decay constant is defined in terms of the element of the axial weak current%
\begin{equation}
\left\langle 0\right\vert A_{\mu }(0)\left\vert B_{c}(q)\right\rangle
=if_{B_{c}}V_{cb}q_{\mu }.
\end{equation}%
The QCD correction factor is%
\begin{equation}
\overline{C}(\alpha _{s})=1-\frac{\alpha _{s}(\mu _{cb}^{2})}{\pi }\left[ 2-%
\frac{m_{b}-m_{c}}{m_{b}+m_{c}}\ln \frac{m_{b}}{m_{c}}\right] .
\end{equation}%
With our parameters $\overline{C}(\alpha _{s})\approx 0.905$ and since this
is rather close to unity, we use it without trying to estimate the higher
order terms.

Let us note the convenient relation%
\begin{equation}
f_{B_{c}}^{2}=\frac{27\mu _{cb}}{8\pi \alpha _{s}(4\mu _{cb}^{2})}\frac{%
m_{b}+m_{c}}{M_{B_{c}}}\overline{C}^{2}(\alpha _{s})\Delta E_{\mathrm{HF}},
\end{equation}%
which for our values of the parameters yields%
\begin{equation}
f_{B_{c}}=65.2\sqrt{\frac{6199}{M_{B_{c}}}}\sqrt{\Delta E_{\mathrm{HF}}},
\end{equation}%
where all the parameters are in suitable powers of $\mathrm{MeV}.$\ \ \ \ \

\subsection{Production Through $Z^{0}$ Decay}

We compute first the width of squarkonium $J^{PC}=0^{++}.$ The $0^{++}$ \
state decays almost entirely into two gluons and $q\overline{q}$ pairs.
Explicit calculations give [89]%
\begin{equation}
\Gamma (0^{++}\rightarrow gg)=\left( \frac{16\pi \alpha _{s}(m^{2})}{3m}%
\right) ^{2}\left\vert \psi (0)\right\vert ^{2},
\end{equation}%
with $m=m_{q}$ $(m_{\widetilde{q}})$ is the constituent mass of quark
(squark) of the bound state system and%
\begin{equation}
\Gamma (0^{++}\rightarrow q\overline{q})=\left( \frac{512\pi \alpha
_{s}(m^{2})}{9m^{2}}\right) \frac{4R^{2}}{\left( 1+4R^{2}\right) ^{2}}%
\left\vert \psi (0)\right\vert ^{2},
\end{equation}%
with $R=\widetilde{m}/m,$ $\widetilde{m}$ $=m_{g}$ $(m_{\widetilde{g}})$ is
the gluon (gluino) mass. In conventional quarkonium, one can extract the
value of $\left\vert \psi (0)\right\vert ^{2}$ from experimental data on the
leptonic width of the quarkonium state (except, of course, for toponium). We
choose to use a tentative input for $\left\vert \psi (0)\right\vert ^{2}$
obtained by a Coulomb like potential ($1/r$ gluonic behaviour), see Eq. (3).
This is justified, since as the mass of the constituents of the bound state
system goes higher (high mass regime), the short-range forces should
approach the Coulomb like interaction whereas the confining linear potential
should be negligible in the short distance limit. We shall stick to a
coulombic wavefunction, namely, [82-84]%
\begin{equation}
\left\vert \psi (0)\right\vert ^{2}=\frac{1}{\pi }\left( \frac{m\alpha
_{s}(m^{2})}{3}\right) ^{3}.
\end{equation}%
The relative branching ratio of Eqs. (39) and (40) is%
\begin{equation}
B=\frac{\Gamma (0^{++}\rightarrow gg)}{\Gamma (0^{++}\rightarrow q\overline{q%
})}.
\end{equation}%
Grifols and M\'{e}ndez [89] computed $\Gamma (0^{++}\rightarrow \widetilde{g}%
\widetilde{g})\approx 0.6$ $\mathrm{MeV}$ and $B>3/8$ for $\alpha _{s}(m_{%
\widetilde{q}}^{2})\approx 0.1609,$ $m_{\widetilde{q}}\approx \mathrm{65}$ $%
\mathrm{GeV}$ and $\widetilde{m}=m_{\widetilde{g}}\geq \mathrm{87}$ $\mathrm{%
GeV}$ or $\widetilde{m}\leq \mathrm{12}$ $\mathrm{GeV.}$

\section{Observed ordinary quark and predicted exotic squark spectroscopy}

The levels are labeled by $S,P,D,$ corresponding to relative orbital angular
momentum $L=0,1,2$ between quark and antiquark. (No candidates for $L\geq 3$
states have been seen yet.) The spin of the quark and antiquark can couple
to either $S=0$ (spin singlet) or $S=1$ (spin triplet) states. The parity of
quark-antiquark or squark-anti-squark states with orbital angular momentum $%
L $ is $P=(-1)^{L+1},$ the charge-conjugation is $C=(-1)^{L+S}.$ Thus, $L=0$
states can be $^{1}S_{0}$ or $^{3}S_{1};$ $L=1$ states can be $^{1}P_{1}$ or
$^{3}P_{0,1,2};$ $L=2$ states can be $^{1}D_{2}$ or $^{3}D_{1,2,3}.$ The
radial quantum number is denoted by $n$ [55]$.$

The experimentally clear spectrum of relatively narrow states below the
open-charm $DD$ threshold of $\mathrm{3730}$ $\mathrm{MeV}$ can be
identified with the $\mathrm{1S,}$ $\mathrm{1P}$ and $\mathrm{2S}$ $c%
\overline{c}$ levels predicted by potential models, which incorporates a
color Coulomb term at short distances and a linear confining term at large
distances [50,51].

A recent interest in charmonium spectroscopy [90] is revived because of the
recent discovery of the long missing $\eta _{c}^{\prime }(\mathrm{2}^{1}%
\mathrm{S}_{0})$ state of binding mass $\mathrm{3638}\pm \mathrm{4}$ $%
\mathrm{MeV}$ by the Belle Collaboration [91,92], which has since then been
confirmed by BABAR [93] and has also been observed by CLEO in $\gamma \gamma
$ collisions [94]. The observation of the $\mathrm{2}^{3}\mathrm{P}_{2}$
state with binding mass $\mathrm{3929\pm 5}$ $\mathrm{MeV.}$ The reported $%
\mathrm{1}^{1}\mathrm{P}_{1}$ $h_{c}$ signal is in the decay chain $\psi
^{\prime }\rightarrow \pi ^{0}h_{c},$ $h_{c}\rightarrow \gamma \eta _{c}.$
The masses found in two different inclusive analysis were $\mathrm{3524.8\pm
0.7}$ $\mathrm{MeV}$ and $\mathrm{3524.4\pm 0.9}$ $\mathrm{MeV}$ (with an
estimated systematic error of $\thicksim $ $\mathrm{1}$ $\mathrm{MeV}$) in
exclusive decay (with six different identified $\eta _{c}$ final states.)
Additional interest in $c\overline{c}$ spectroscopy has followed the
discovery of the remarkable $X(\mathrm{3872})$ by Belle [95] and CDF [96] in
$B$ decays to $J/\psi \pi ^{+}\pi ^{-\text{ }};$ assuming that this is a
real resonance rather than a threshold effect, the $X(\mathrm{3872})$ is
presumably either a $DD^{\ast }$ charmed meson molecules [97-99] or a narrow
$J=2$ $D$-wave $c\overline{c}$ state [100,101]. Very recent observations of
the $X(\mathrm{3872})$ in $\gamma J/\psi $ and $\omega J/\psi $ by Belle
support $1^{++}$ $DD^{\ast }$ molecule assignment [102,103]. Experimental
activity in the spin-singlet $P$-wave, with recent reports of the
observation of exclusive $1^{1}P_{J}$ $h_{c}$ state by CLEO measurement of
mass $\mathrm{3524.65\pm 0.55}$ $\mathrm{MeV}$ [104,105]. The $\psi (\mathrm{%
3770})$ is generally assumed to be $^{3}\mathrm{D}_{1}$ $c\overline{c}$
state, perhaps with a significant $\mathrm{2}^{3}\mathrm{S}_{1}$ component
[106]. The four known $c\overline{c}$ states above the $DD$ threshold, \ $%
\psi (\mathrm{3770}),\psi (\mathrm{4040}),\psi (\mathrm{4159})$ and $\psi (%
\mathrm{4415})$ are of special interest because they are easily produced at $%
e^{+}e^{-}$ machines. The mass $\psi (\mathrm{4040})$ which is a very
interesting case for the study of strong decays. The $1^{--}$ $\psi (\mathrm{%
4159})$ is $\mathrm{2}^{3}\mathrm{D}_{1}$ $c\overline{c}$ assignment. The
final known above $DD$ threshold is the $1^{--}$ $\psi (\mathrm{4415})$ has
the assignment $\mathrm{4}^{3}\mathrm{S}_{1}.$

On the other hand, the searched pseudoscalar bottomonium $\eta _{b}(\mathrm{1%
}^{1}\mathrm{S}_{0})$ meson, the new observed meson $\Upsilon (\mathrm{4}^{3}%
\mathrm{S}_{1})=\mathrm{10579.4\pm 1.2}$ $\mathrm{MeV}$ and the $\mathrm{1}%
^{3}\mathrm{D}_{J}$ (probably all or mostly $J=2$) state with bound mass $%
\mathrm{10161.1\pm 1.7}$ $\mathrm{MeV}$ [55] and the pseudoscalar charmed
bottom meson with an averaged experimental mass $B_{c}(0^{-+})=\mathrm{%
6286\pm 5}$ $\ \mathrm{MeV}$ by Yao \textit{et al.} have also revived this
interest [55].

We first apply this model described above to the ordinary quarkonia known
through the SM. Consequently, we calculate the $c\overline{c},b\overline{b}$
and $c\overline{b}(b\overline{c})$ quarkonium binding mass spectra in close
agreement with above up-to-date experimental findings. The theoretically
calculated quarkonium binding masses together with their $\mathrm{S}$-states
hyperfine splittings are listed in Tables 1-2. The hyperfine mass splittings
of the $\mathrm{1S}$ $c\overline{b}$ state is predicted by the present
potential model and other models are listed in Table 3. The calculated
binding masses are found to be in close agreement to the recently observed
ones. This allows us to extend this study to the unknown spectra of the
squarks to predict their binding masses in a unified way. Further, in Tables
1-3, all hyperfine splitting calculations of the potential model try to
reproduce the old experimental values, while the lattice calculations and
perturbative QCD favor the new values. No confirmed experimental data to
check these predictions are available yet. The experimental finding $\Delta
_{\mathrm{HF}}(\mathrm{2S,}c\overline{c})=\psi ^{\prime }(\mathrm{2}^{3}%
\mathrm{S}_{1})-\eta _{c}^{\prime }(\mathrm{2}^{1}\mathrm{S}%
_{0})\thickapprox \mathrm{48.093\pm 3.97}$ $\mathrm{MeV}$ is found to be
lower than the calculated values from the potential models [19,26]. In all
cases, where comparison with the other models are significantly smaller than
the splittings found by Eichten and Quigg [71] and Gupta and Johnson [107].
The QCD sum rules [108,109] finds the hyperfine splitting of the bottomonium
$\Delta _{\mathrm{HF}}(\mathrm{1S,}b\overline{b},$theory)$=63_{-51}^{+29}$ $%
\mathrm{MeV}$ with the central value agrees well to several $\mathrm{MeV}$
with expectation. However, the uncertainty is too large to distinguish
between the potential models. A lattice calculation [110] gives the
hyperfine splitting $\Delta _{\mathrm{HF}}(\mathrm{1S,}b\overline{b},$theory)%
$\mathrm{=60}$ $\mathrm{MeV}$ with a large uncertainty. The central value
seems to be close to our model, but the uncertainty is big enough to be
consistent with all the potential models quoted here. Furthermore, our model
predicts nearly an approximate hyperfine splitting for the $\mathrm{1S}$
bottomonium and $\mathrm{2S}$ charmonium as in the other potential models
[34,35,49,71], lattice [111,112] and perturbation QCD [113,114].

The level fine and hyperfine splittings in charmonium and bottomonium
together with the experimental and other models findings are listed in Table
4. The fine splitting in charmonium is found to be $M_{\psi ^{\prime }(%
\mathrm{2S})}-M_{J/\psi (\mathrm{1S})}\mathrm{=597}$ $\mathrm{MeV.}$ It is
within $\mathrm{7.8}$ $\mathrm{MeV}$ from the experimental value. However,
the splitting for bottomonium is found to be $M_{\Upsilon ^{\prime }(\mathrm{%
2S})}-M_{\Upsilon (\mathrm{1S})}\mathrm{=571}$ $\mathrm{MeV.}$ It is within $%
\mathrm{8}$ $\mathrm{MeV}$ from the experimental value.

Motivated by the great success of our earlier applications [33,38-49], we
extend this study to produce the binding masses of the sbottom with
anti-sbottom and heavy quark with anti-sbottom. In Table 5, we show the
results about $\widetilde{b}\overline{\widetilde{b}}$ states. The numerical
results about ($q\overline{\widetilde{b})},(q=b,c)$ states are shown in
Table 6 and Table 7, respectively. Because sbottom is a spin zero particle,
the spectrum of the corresponding bound states is simpler than that of $(q%
\overline{q})$ states. We can see from these tables that when $m_{\widetilde{%
b}}\thickapprox m_{b},$ the binding masses of the lowest states ($b\overline{%
b}),(b\widetilde{\overline{b}})$ and $\widetilde{b}\overline{\widetilde{b}}$
are very close to each other. This is reasonable because the strong
interaction for these states is similar. Also, the leptonic decay widths of
the sbottomonium system resonances for different values of the resonance
sbottom mass and $e_{q}=\frac{1}{3}$are shown in Table 8.

In this analysis, taking the unknown mass of the sbottom quark close to the
ordinary bottom quark in the $\widetilde{b}\overline{\widetilde{b}}$ pair
with $m_{\widetilde{b}}=\mathrm{4.5}$ $\mathrm{MeV,}$ we find out the fine
splittings for the $\mathrm{S}$-states as follows: $\mathrm{569,335,252}$ $%
\mathrm{MeV}$ and for $m_{\widetilde{b}}=\mathrm{5.0}$ $\mathrm{MeV}$ as: $%
\mathrm{571,332,249}$ $\mathrm{MeV.}$ Thus, we remark that the fine
splittings of the ordinary $c\overline{c},b\overline{b}$ and $c\overline{b}(b%
\overline{c})$ quarkonium systems and the ($q\overline{\widetilde{b}}%
),(q=b,c)$ and $\widetilde{b}\overline{\widetilde{b}}$ exotic quarkonium
states are nearly same [130]. Finally, in general, the potential models seem
to reproduce the experimental values much better. This feature would be
understandable, since the potential models contain much more input
parameters than the lattice or perturbative QCD models.

We point out that Chang \textit{et al}. [78] used a relativistic model to
calculate part of these bound states for some $%
J^{PC}=0^{++},1^{--},2^{++},3^{--}$ corresponding to $%
n^{3}P_{0},n^{3}D_{1},n^{3}P_{2}$ and $n^{3}D_{3},$ $n=0,1,2,3,$
respectively, which is entirely different than our c.o.g. calculations.
Calculating states like $n^{3}P_{1},n^{3}P_{2}$ and $n^{3}D_{2}$ could help
us to compare with other models [130].

\section{Conclusions}

We have obtained the bound state masses and hyperfine energy splittings of a
flavor-independent static quarkonium potential model for few ordinary
quarkonium and scalar squarkonium mesons within the framework of the shifted
$N$-expansion technique for $L=0,1,2$ states (see Tables 1-7). In Tables
1-4, we have shown the effectiveness of the employed static quarkonium
potential model in producing the quarkonium bound-state masses to several $%
\mathrm{MeV}$. Encouraged by this success of a flavor-independent potential
model, we have also predicted the bound state masses of few unknown
squarkonium systems for low- to- high squark masses ($\mathrm{3.0}$ $\mathrm{%
GeV}$-$\mathrm{150.0}$ $\mathrm{GeV}$) as shown in Tables 5-7. In finding
the unknown squarkonium energy splittings, we have used the quarkonium
strong coupling constant $\alpha _{s}(m^{2})$ values to predict the
squarkonium energy splittings. Essentially, this is because the type of
interaction between two squarks is very similar to the interaction between
two quarks (for reviews see, for example, [36] and references therein). Such
an interaction is part of the one gluon exchange interaction and is
responsible for the mass differences. Apparently, the squarkonium fine and
hyperfine splittings are found to be nearly same as their quarkonium
counterparts if the squark mass is chosen near the quark mass (i.e., $m_{%
\widetilde{q}}\simeq m_{q}$). Our conclusions are also consistent with the
conclusions made by Ref. [130]. This is reasonable because the strong
interaction for these states is similar (see, for example, [130]). The
exotic bound states are likely to form bound states in an entirely similar
fashion as ordinary quarks form bound state, i.e., quarkonium [89]. Since
the same potential model is used for conventional QCD bound states and
sbottomonium, so it is obvious, from the present work and Ref. [130], that
the mechanism responsible for the binding of a $b\overline{b}$ couple is the
same one responsible for the binding a scalar $\widetilde{b}\overline{%
\widetilde{b}}.$ In general, the type of interaction in squarkonium is very
similar to that in quarkonium [36].

Furthermore, the calculation of the leptonic decay constant is important in
predicting the cross-section. Therefore, we have calculated the leptonic
decay constants in Table 8. We conclude that sbottomonium bound states can
be detected as resonances at LHC. Above quark masses of about $\mathrm{2}$ $%
\mathrm{GeV,}$ $\Gamma _{e}$ is much higher than the experimental upper
limits even for squarks of charge $\frac{2}{3}.$ Moreover, $\Gamma _{e}$
increases rapidly with the squark mass. In addition, in the supersymmetric
front, squarks, sleptons and gauginos do also have a probability to be
pairwise produced at LHC [89]. Squarks are likely to form bound states in an
entirely similar fashion as conventional quarks from bound state, i.e.,
quarkonium [89]. The model-dependent squark leptonic decay widths are bigger
than the ordinary quarks for $m_{\widetilde{b}}<m_{b}$ and $m_{\widetilde{b}%
}>\mathrm{10}$ $\mathrm{GeV}$ and give decay smaller than the energy
splitting between bound states which is assumed to be a narrow resonance. \
As an illustrative example, with given squark mass $m_{\widetilde{q}}=%
\mathrm{65}$ $\mathrm{GeV}$ and strong coupling constant $\alpha _{s}(m_{%
\widetilde{q}}^{2})=0.1609$ in Ref. [89]$,$ we solve the Schr\"{o}dinger
equation for the strictly phenomenological scalar potential (3) to obtain
the squarkonium binding masses $M(1^{3}S_{1})=\mathrm{127.745}$ $\mathrm{GeV}
$ and $M(1^{1}S_{0})=\mathrm{127.577}$ $\mathrm{GeV}$ for the singlet and
triplet ground states, respectively. We, further, calculate the hyperfine
splitting energy $\Delta E=\mathrm{167.72}$ $\mathrm{MeV}$ and decay
constant $\Gamma _{e}=\mathrm{0.556}$ $\mathrm{KeV}$ which is smaller than
the energy-beam width$\mathrm{.}$ Hence, we conclude that this decay width
is a wide resonance. On the other hand, the hidden supersymmetry showing up
the squarkonium production through the $Z^{0}$ decay into a photon and the
lowest lying state of the squarkonium system, the $J^{PC}=0^{++}$ state ($%
^{1}S_{0}$ in spectroscopic notation). The integrated area of the resonance
cross-section, Eq. (32), has a large (small) value for light (heavy) squark
mass, respectively. For example, from Table 8, we find $0.008\times
10^{-6}GeV^{-1}\leq \int \sigma _{res}dE\leq 2.56\times 10^{-6}$ $GeV^{-1}$
for the range $\mathrm{3.0}$ $\mathrm{GeV}\leq m_{\widetilde{b}}\leq \mathrm{%
150.0}$ $\mathrm{GeV}$ sbottom quark mass resonance. Hence, the detection of
a low-mass squarks at LHC are much favored.

Meanwhile, from the width of squarkonium $0^{++},$ Eqs. (39)-(42), we see
that the relative branching ratio, $B=0.662$ for $m_{\widetilde{b}}=\mathrm{%
60}$ $\mathrm{GeV}$ and $m_{\widetilde{g}}=\mathrm{87}$ $\mathrm{GeV}$ by
using the Coulomb potential result for $\left\vert \psi (0)\right\vert ^{2}$
we get $\Gamma (0^{++}\rightarrow \widetilde{g}\widetilde{g})=\mathrm{2.54}$
$\mathrm{MeV}$ with the strong coupling constant is taken relatively larger
than Ref. [89]$\mathrm{.}$ For low-mass case $m_{\widetilde{b}}=\mathrm{10}$
$\mathrm{GeV}$ and $m_{\widetilde{g}}=\mathrm{12}$ $\mathrm{GeV,}$ we obtain
$B=0.498\approx 4/8$ which is above $3/8$ and $\Gamma (0^{++}\rightarrow
\widetilde{g}\widetilde{g})=\mathrm{0.424}$ $\mathrm{MeV}$ which is
consistent with [89]$\mathrm{.}$ This result also favors low bottom squark
mass [16].

\acknowledgments The authors thank the anonymous kind referee(s) for the
constructive comments and suggestions that have improved the paper greatly.
They are also grateful for the partial support provided by the Scientific
and Technological Research Council of Turkey (T\"{U}B\.{I}TAK).

\bigskip

\bigskip \baselineskip= 2\baselineskip% double space the text
%\end{document}
\bigskip

\begin{table}[tbp]
\caption{Observed and calculated binding mass spectrum of $c\overline{c}$
states (in $\mathrm{MeV}$). $\Delta X$ denotes the mass shift of the
spin-singlet state ($n^{1}S_{0}$) from the spin-triplet state ($n^{3}S_{1}$%
). }
\label{Table 1}%
\begin{tabular}{lllllllll}
State & $n(J^{PC})$ & Our work & NR [59] & GI [60] & EQ [71] & GJ [107] & MZ
[34,35] & PDG [55] \\
\tableline$J/\psi (1^{3}S_{1})$ & $1(1^{--})$ & $3097$ & $3090$ & $3098$ & $%
3097$ & $3097$ & $3097$ & $3096.916\pm 0.011$ \\
$\Delta 1^{1}S_{0}$ & $1(0^{-+})$ & $-117$ & $-108$ & $-123$ & $-117$ & $%
-117 $ & $-117$ & $-116.516\pm 1.189$ \\
$1P$ (c.o.g) & $1(0,1,2)^{++}$ & $3521$ & $3524.3$ & $3525$ & $3492$ & $3526$
& $3521$ & $3525.3\pm 0.11$ \\
$\psi ^{\prime }(2^{3}S_{1})$ & $2(1^{--})$ & $3694$ & $3672$ & $3676$ & $%
3686$ & $3685$ & $3690$ & $3686.093\pm 0.034$ \\
$\Delta 2^{1}S_{0}$ & $2(0^{-+})$ & $-65.9$ & $-42$ & $-53$ & $-78$ & $-68$
& $-72$ & $-48.093\pm 3.966$ \\
$1D$ (c.o.g) \tablenotemark[1]\tablenotetext[1]{$1^{3}D_{1}$ state.} & $%
1(1,2,3)^{--}$ & $3806$ & $3801$ & $3842$ & $-$ & $-$ & $-$ & $3771.1\pm 2.4$
\\
$2P$ (c.o.g) \tablenotemark[2]\tablenotetext[2]{$2^{3}P_{2}$ state.} & $%
2(0,1,2)^{++}$ & $3944$ & $3943$ & $3963.3$ & $-$ & $-$ & $-$ & $3929\pm 5$
\\
$\psi (3^{3}S_{1})$ & $3(1^{--})$ & $4078$ & $4072$ & $4100$ & $-$ & $-$ & $%
- $ & 4040$\pm 10$ \\
$2D$ (c.o.g) & $2(1,2,3)^{--}$ & $4150$ & $4161.2$ & $4211.4$ & $-$ & $-$ & $%
-$ & 4159$\pm 20$ \\
$3P$ (c.o.g) & $3(0,1,2)^{++}$ & $4265$ & $4288.9$ & $4325.3$ & $-$ & $-$ & $%
-$ & $-$ \\
$\psi (4^{3}S_{1})$ & $4(1^{--})$ & $4377$ & $4406$ & $4450$ & $-$ & $-$ & $%
- $ & $4415\pm 6$ \\
$3D$ (c.o.g) & $3(1,2,3)^{--}$ & $4430$ & $-$ & $-$ & $-$ & $-$ & $-$ & $-$
\\
$\psi (5^{3}S_{1})$ & $5(1^{--})$ & $4628$ & $-$ & $-$ & $-$ & $-$ & $-$ & $%
- $%
\end{tabular}%
\end{table}

\bigskip

\begin{table}[tbp]
\caption{Observed and calculated binding mass spectrum of $b\overline{b}$
states (in $\mathrm{MeV}$). $\Delta X$ denotes the mass shift of the
spin-singlet state ($n^{1}S_{0}$) from the spin-triplet state ($n^{3}S_{1}$%
). }
\label{Table 2}%
\begin{tabular}{lllllll}
State & $n(J^{PC})$ & Our work & EQ [71] & KR [115] & MZ [34,35] & PDG [55]
\\
\tableline$\Upsilon (1^{3}S_{1})$ & $1(1^{--})$ & $9460$ & $9464$ & $-$ & $%
9460$ & $9460.30\pm 0.26$ \\
$\Delta 1^{1}S_{0}$ & $1(0^{-+})$ & $-57.9$ & $-87$ & $-$ & $-56.7$ & $%
(160)? $ \\
$1P$(c.o.g) & $1(0,1,2)^{++}$ & $9900$ & $9873$ & $9903$ & $9900$ & $%
9899.87\pm 0.41$ \\
$\Upsilon (2^{3}S_{1})$ & $2(1^{--})$ & $10031$ & $10007$ & $-$ & $10023$ & $%
10023.26\pm 0.31$ \\
$\Delta 2^{1}S_{0}$ & $2(0^{-+})$ & $-23.2$ & $-44$ & $-$ & $-28$ & $-$ \\
$1D$(c.o.g) $(1^{3}D_{J}$ state$)$\tablenotemark[1]%
\tablenotetext[1]{Probably all or mostly $J=2$.} & $1(1,2,3)^{--}$ & $10155$
& $10127$ & $10156$ & $10155$ & $10161.1\pm 1.7$ \\
$2P$(c.o.g) & $2(0,1,2)^{++}$ & $10261$ & $10231$ & $10259$ & $10260$ & $%
10260.237\pm 0.56$ \\
$\Upsilon (3^{3}S_{1})$ & $3(1^{--})$ & $10364$ & 10339 & $-$ & $10355$ & $%
10355.2\pm 0.5$ \\
$2D$(c.o.g) & $2(1,2,3)^{--}$ & $10438$ & $-$ & $10441$ & $10438$ & $-$ \\
$3P$(c.o.g) & $3(0,1,2)^{++}$ & $10527$ & $-$ & $10520$ & $10525$ & $-$ \\
$\Upsilon (4^{3}S_{1})$ & $4(1^{--})$ & $10614$ & $-$ & $-$ & $-$ & $%
10579.4\pm 1.2$ \\
$3D$(c.o.g) & $3(1,2,3)^{--}$ & $10666$ & $-$ & $-$ & $-$ & $-$ \\
$\Upsilon (5^{3}S_{1})$ & $5(1^{--})$ & $10820$ & $-$ & $-$ & $-$ & $-$%
\end{tabular}%
\end{table}

\bigskip \bigskip \bigskip\
\begin{table}[tbp]
\caption{The calculated $\overline{b}c$ binding masses of the lowest $%
\mathrm{S}$-states and its splitting compared with the other authors (in $%
\mathrm{MeV}).$}%
\begin{tabular}{llll}
Work & $M_{B_{c}}(1^{1}S_{0})$\tablenotemark[1]\tablenotetext[1]{The
averaged observed mass.} & $M_{B_{c}^{\ast }}(1^{3}S_{1})$ & $\Delta _{1S}$
\\
\tableline PDG (Yao et al.) [55] & $6286\pm 5$ & $-$ & $-$ \\
Our work & $6290.8$ & $6349$ & $58.2$ \\
Motyka and Zalewiski [34,35] & $6291$ & $6349$ & $58$ \\
Eichten and Quigg [71] & $6264$ & $6337$ & $73$ \\
Colangelo and Fazio [117] & $6280$ & $6350$ & $70$ \\
Chabab [109] & $6250\pm 200$ & $-$ & $-$ \\
Baker et al. [118] & $6287$ & $6372$ & $85$ \\
Roncaglia et al. [119] & $6255$ & $6320$ & $65$ \\
Godfrey et al. [120] & $6270$ & $6340$ & $70$ \\
Bagan et al. [121] & $6255\pm 20$ & $6330\pm 20$ & $75$ \\
Brambilla et al. [122] & $-$ & $6326_{-9}^{+29}$ & $60$ \\
Baldicchi and Prosperi [85-87] & $6194\sim 6292$ & $6284\sim 6357$ & $65\leq
\Delta _{1S}\leq 90$ \\
SLET [49]\tablenotemark[2]\tablenotetext[2]{Averaging over the five values
in Table 1 of [41].} & $6253_{-6}^{+13}$ & $6328_{-9}^{+7}$ & $68\leq \Delta
_{1S}\leq 83$ \\
SLET [49]\tablenotemark[3]\tablenotetext[3]{We treat results of [71] in the
same manner.} & $6258_{-11}^{+8}$ & $6333_{-14}^{+2}$ & $-$ \\
Chen and Kuang [61,62] & $6310$ & $6355$ & $45$ \\
Gershtein et al. [123] & $6253$ & $6317$ & $64$ \\
Gupta and Johnson [107] & $6267$ & $6308$ & $41$%
\end{tabular}%
\end{table}

\mediumtext

\bigskip \bigskip \bigskip
\begin{table}[tbp]
\caption{Level hyperfine and fine splittings in charmonium and bottomonium
(in $MeV)$.}
\label{table1 3}%
\begin{tabular}{llllllllll}
Level splitting\tablenotetext[1]{Potential model.} & Our work & [71]%
\tablenotemark[1] & [34,35]\tablenotemark[1] & [124-127]\tablenotemark[1] &
[128]\tablenotemark[1] & [111]\tablenotemark[2] & [129]\tablenotemark[2] &
[113,114]\tablenotemark[3] & PDG [55] \\
\tableline$\Delta _{HF}^{(c\overline{c})}(2S)=M_{\psi ^{\prime
}(2S)}-M_{\eta _{c}^{\prime }(2S)}$ & $66$ & $78$ & $72$ & $98$ & $92$ & $43$
& $-$ & $38$ & $48.093\pm 3.966$ \\
$\Delta _{HF}^{(b\overline{b})}(1S)=M_{\Upsilon (1S)}-M_{\eta _{b}(1S)}$%
\tablenotetext[2]{Lattice.} & $58$ & $87$ & $57$ & $60$ & $45$ & $-$ & $51$
& $44$ & $(160)?$ \\
$\Delta _{HF}^{(b\overline{b})}(2S)=M_{\Upsilon ^{\prime }(2S)}-M_{\eta
_{b}^{\prime }(2S)}$\tablenotetext[3]{Perturbative QCD.} & $23$ & $44$ & $28$
& $30$ & $28$ & $-$ & $-$ & $21$ & $-$ \\
$\Delta _{F}^{(c\overline{c})}=M_{\psi ^{\prime }(2S)}-M_{J/\psi (1S)}$ & $%
597$ &  &  &  &  &  &  &  & $589.177\pm 0.023$ \\
$\Delta _{F}^{(c\overline{c})}=M_{\psi ^{\prime }(3S)}-M_{\psi ^{\prime
}(2S)}$ & $384$ &  &  &  &  &  &  &  & $353.907\pm 9.966$ \\
$\Delta _{F}^{(b\overline{b})}=M_{\Upsilon ^{\prime }(2S)}-M_{\Upsilon (1S)}$
& $571$ &  &  &  &  &  &  &  & $562.96\pm 0.05$ \\
$\Delta _{F}^{(b\overline{b})}=M_{\Upsilon ^{\prime }(3S)}-M_{\Upsilon
^{\prime }(2S)}$ & $333$ &  &  &  &  &  &  &  & $331.94\pm 0.19$ \\
$\Delta _{F}^{(b\overline{b})}=M_{\Upsilon ^{\prime }(4S)}-M_{\Upsilon
^{\prime }(3S)}$ & $250$ &  &  &  &  &  &  &  & $224.2\pm 0.7$%
\end{tabular}%
\end{table}

\bigskip
\begin{table}[tbp]
\caption{The calculated binding energy masses of $\widetilde{b}\overline{%
\widetilde{b}}$ pair (in $MeV).$}
\label{table5}%
\begin{tabular}{lllllllll}
State & $m_{\widetilde{b}}$\tablenotemark[1]\tablenotetext[1]{The mass is in
$GeV$.} & $nS$ & $nP$\tablenotemark[2]\tablenotetext[2]{The calculated
c.o.g. binding energy mass for states $n(0)^{++},n(1)^{++}$ and $n(2)^{++}$.}
& $nD$\tablenotemark[3]\tablenotetext[3]{The calculated c.o.g. binding
energy mass for states $n(1)^{--},n(2)^{--}$ and $n(3)^{--}$.} & $m_{%
\widetilde{b}}$\tablenotemark[1] & $nS$ & $nP$ & $nD$ \\
\tableline$n=1$ & $3.0$ & $6033$ & $6455$ & $6718$ & $3.5$ & $6976$ & $7402$
& $7661$ \\
$n=2$ &  & $6601$ & $6835$ & $7021$ &  & $7543$ & $7775$ & $7958$ \\
$n=3$ &  & $6950$ & $7119$ & $7267$ &  & $7886$ & $8053$ & $8198$ \\
$n=4$ &  & $7216$ & $7352$ & $7477$ &  & $8146$ & $8280$ & $8402$ \\
$n=1$ & $4.0$ & $7925$ & $8356$ & $8613$ & $4.5$ & $8880$ & $9316$ & $9572$
\\
$n=2$ &  & $8493$ & $8724$ & $8904$ &  & $9449$ & $9679$ & $9858$ \\
$n=3$ &  & $8831$ & $8997$ & $9139$ &  & $9784$ & $9948$ & $10088$ \\
$n=4$ &  & $9087$ & $9219$ & $9339$ &  & $10036$ & $10167$ & $10284$ \\
$n=1$ & $5.0$ & $9839$ & $10280$ & $10535$ & $5.5$ & $10800$ & $11248$ & $%
11501$ \\
$n=2$ &  & $10410$ & $10640$ & $10817$ &  & $11375$ & $11604$ & $11780$ \\
$n=3$ &  & $10742$ & $10905$ & $11043$ &  & $11705$ & $11866$ & $12003$ \\
$n=4$ &  & $10991$ & $11120$ & $11236$ &  & $11951$ & $12079$ & $12193$ \\
$n=1$ & $6.0$ & $11764$ & $12219$ & $12471$ & $8.0$ & $15639$ & $16121$ & $%
16373$ \\
$n=2$ &  & $12343$ & $12572$ & $12747$ &  & $16236$ & $16468$ & $16639$ \\
$n=3$ &  & $12671$ & $12831$ & $12967$ &  & $16560$ & $16719$ & $16851$ \\
$n=4$ &  & $12915$ & $13041$ & $13155$ &  & $16797$ & $16921$ & $17031$ \\
$n=1$ & $10.0$ & $19533$ & $20044$ & $20298$ & $20.0$ & $39120$ & $39793$ & $%
40066$ \\
$n=2$ &  & $20152$ & $20388$ & $20558$ &  & $39879$ & $40142$ & $40314$ \\
$n=3$ &  & $20476$ & $20634$ & $20764$ &  & $40215$ & $40379$ & $40506$ \\
$n=4$ &  & $20709$ & $20831$ & $20938$ &  & $40442$ & $40563$ & $40666$ \\
$n=1$ & $40.0$ & $78472$ & $79496$ & $79821$ & $60.0$ & $117887$ & $119278$
& $119663$ \\
$n=2$ &  & $79559$ & $79884$ & $80072$ &  & $119329$ & $119717$ & $119925$
\\
$n=3$ &  & $79942$ & $80126$ & $80259$ &  & $119766$ & $119973$ & $120114$
\\
$n=4$ &  & $80179$ & $80307$ & $80410$ &  & $120019$ & $120157$ & $120264$%
\end{tabular}%
\end{table}

\bigskip

\begin{table}[tbp]
\caption{The calculated binding energy masses of $c\overline{\widetilde{b}}$
pair (in $MeV).$}
\label{table6}%
\begin{tabular}{lllllllll}
State & $m_{\widetilde{b}}$\tablenotemark[1]\tablenotetext[1]{The mass is in
$GeV$.} & $nS$ & $nP$ & $nD$ & $m_{\widetilde{b}}$\tablenotemark[1] & $nS$ &
$nP$ & $nD$ \\
\tableline$n=1$ & $4.0$ & $5562$ & $5982$ & $6254$ & $4.5$ & $6052$ & $6471$
& $6743$ \\
$n=2$ &  & $6140$ & $6381$ & $6576$ &  & $6629$ & $6869$ & $7063$ \\
$n=3$ &  & $6504$ & $6682$ & $6837$ &  & $6992$ & $7168$ & $7323$ \\
$n=4$ &  & $6785$ & $6929$ & $7061$ &  & $7271$ & $7415$ & $7546$ \\
$n=1$ & $4.8$ & $6346$ & $6766$ & $7037$ & $4.9$ & $6445$ & $6864$ & $7135$
\\
$n=2$ &  & $6923$ & $7162$ & $7356$ &  & $7021$ & $7261$ & $7454$ \\
$n=3$ &  & $7285$ & $7461$ & $7616$ &  & $7383$ & $7559$ & $7714$ \\
$n=4$ &  & $7564$ & $7707$ & $7838$ &  & $7662$ & $7805$ & $7935$ \\
$n=1$ & $5.0$ & $6543$ & $6963$ & $7233$ & $5.5$ & 7036 & $7456$ & $7726$ \\
$n=2$ &  & $7119$ & $7359$ & $7552$ &  & 7611 & $7851$ & $8044$ \\
$n=3$ &  & $7481$ & $7657$ & $7812$ &  & 7972 & $8148$ & $8302$ \\
$n=4$ &  & $7760$ & $7903$ & $8033$ &  & 8250 & $8393$ & $8523$ \\
$n=1$ & $6.0$ & $7530$ & $7950$ & $8219$ & $8.0$ & $9513$ & $9933$ & $10201$
\\
$n=2$ &  & $8105$ & $8344$ & $8536$ &  & $10086$ & $10324$ & $10515$ \\
$n=3$ &  & $8465$ & $8640$ & $8794$ &  & $10444$ & $10618$ & $10771$ \\
$n=4$ &  & $8742$ & $8884$ & $9014$ &  & $10719$ & $10860$ & $10989$ \\
$n=1$ & $10.0$ & $11502$ & $11922$ & $12189$ & $20.0$ & $21479$ & $21900$ & $%
22166$ \\
$n=2$ &  & $12074$ & $12312$ & $12503$ &  & $22050$ & $22286$ & $22476$ \\
$n=3$ &  & $12431$ & $12605$ & $12757$ &  & $22404$ & $22577$ & $22727$ \\
$n=4$ &  & $12705$ & $12846$ & $12974$ &  & $22676$ & $22815$ & $22942$ \\
$n=1$ & $40.0$ & $41467$ & $41888$ & $42153$ & $60.0$ & $61463$ & $61884$ & $%
62149$ \\
$n=2$ &  & $42037$ & $42273$ & $42462$ &  & $62033$ & $62269$ & $62457$ \\
$n=3$ &  & $42390$ & $42562$ & $42712$ &  & $62385$ & $62557$ & $62706$ \\
$n=4$ &  & $42660$ & $42799$ & $42925$ &  & $62655$ & $62794$ & $62920$%
\end{tabular}%
\end{table}

\bigskip

\begin{table}[tbp]
\caption{The calculated binding energy masses of $b\overline{\widetilde{b}}$
pair (in $MeV).$}
\label{table7}%
\begin{tabular}{lllllllll}
State & $m_{\widetilde{b}}$\tablenotemark[1]\tablenotetext[1]{The mass is in
$GeV$.} & $nS$ & $nP$ & $nD$ & $m_{\widetilde{b}}$\tablenotemark[1] & $nS$ &
$nP$ & $nD$ \\
\tableline$n=1$ & $4.0$ & $8695$ & $9130$ & $9385$ & $4.5$ & $9170$ & $9608$
& $9863$ \\
$n=2$ &  & $9263$ & $9494$ & $9673$ &  & $9740$ & $9970$ & $10148$ \\
$n=3$ &  & $9599$ & $9763$ & $9904$ &  & $10074$ & $10238$ & $10377$ \\
$n=4$ &  & $9852$ & $9983$ & $10101$ &  & $10325$ & $10455$ & $10572$ \\
$n=1$ & $4.8$ & $9458$ & $9897$ & $10152$ & $4.9$ & $9554$ & $9994$ & $10248$
\\
$n=2$ &  & $10028$ & $10258$ & $10435$ &  & $10124$ & $10354$ & $10531$ \\
$n=3$ &  & $10361$ & $10524$ & $10663$ &  & $10457$ & $10620$ & $10759$ \\
$n=4$ &  & $10611$ & $10741$ & $10857$ &  & $10707$ & $10836$ & $10953$ \\
$n=1$ & $5.0$ & $9650$ & $10090$ & $10345$ & $5.5$ & $10131$ & $10575$ & $%
10829$ \\
$n=2$ &  & $10221$ & $10450$ & $10627$ &  & $10704$ & $10933$ & $11110$ \\
$n=3$ &  & $10553$ & $10716$ & $10855$ &  & $11035$ & $11197$ & $11336$ \\
$n=4$ &  & $10803$ & $10932$ & $11048$ &  & $11283$ & $11412$ & $11528$ \\
$n=1$ & $6.0$ & $10616$ & $11061$ & $11315$ & $8.0$ & $12567$ & $13022$ & $%
13274$ \\
$n=2$ &  & $11189$ & $11419$ & $11594$ &  & $13145$ & $13375$ & $13550$ \\
$n=3$ &  & $11520$ & $11681$ & $11819$ &  & $13474$ & $13634$ & $13770$ \\
$n=4$ &  & $11767$ & $11895$ & $12010$ &  & $13717$ & $13844$ & $13957$ \\
$n=1$ & $10.0$ & $14534$ & $14995$ & $15247$ & $20.0$ & $24457$ & $24935$ & $%
25187$ \\
$n=2$ &  & $15117$ & $15347$ & $15520$ &  & $25051$ & $25283$ & $25454$ \\
$n=3$ &  & $15444$ & $15603$ & $15738$ &  & $25376$ & $25534$ & $25667$ \\
$n=4$ &  & $15685$ & $15811$ & $15923$ &  & $25613$ & $25737$ & $25847$ \\
$n=1$ & $40.0$ & $44410$ & $44900$ & $45153$ & $60.0$ & $64393$ & $64888$ & $%
65140$ \\
$n=2$ &  & $45013$ & $45246$ & $45417$ &  & $64999$ & $65233$ & $65404$ \\
$n=3$ &  & $45337$ & $45495$ & $45627$ &  & $65323$ & $65481$ & $65613$ \\
$n=4$ &  & $45572$ & $45696$ & $45805$ &  & $65558$ & $65681$ & $65789$%
\end{tabular}%
\end{table}

\bigskip

\begin{table}[tbp]
\caption{Leptonic widths of the sbottomonium system resonances for different
values of the resonance sbottom mass and $e_{q}=\frac{1}{3}.$}%
\begin{tabular}{lllllllll}
State & $m(GeV)$ & $\Gamma _{e}(KeV)$ & $m(GeV)$ & $\Gamma _{e}(KeV)$ & $%
m(GeV)$ & $\Gamma _{e}(KeV)$ & $m(GeV)$ & $\Gamma _{e}(KeV)$ \\
\tableline$n=1$ & $2.0$ & $1.715$ & $2.5$ & $1.635$ & $3.0$ & $1.571$ & $3.5$
& $1.523$ \\
$n=2$ &  & $0.694$ &  & $0.657$ &  & $0.620$ &  & $0.587$ \\
$n=3$ &  & $0.464$ &  & $0.445$ &  & $0.423$ &  & $0.401$ \\
$n=1$ & $4.0$ & $1.487$ & $4.5$ & $1.461$ & $4.8$ & $1.450$ & $4.9$ & $1.447$
\\
$n=2$ &  & $0.557$ &  & $0.530$ &  & $0.516$ &  & $0.512$ \\
$n=3$ &  & $0.381$ &  & $0.362$ &  & $0.352$ &  & $0.349$ \\
$n=1$ & $5.0$ & $1.444$ & $5.5$ & $1.433$ & $10.0$ & $1.504$ & $20.0$ & $%
2.032$ \\
$n=2$ &  & $0.507$ &  & $0.487$ &  & $0.381$ &  & $0.315$ \\
$n=3$ &  & $0.345$ &  & $0.330$ &  & $0.244$ &  & $0.174$ \\
$n=1$ & $40.0$ & $3.405$ & $60.0$ & $4.869$ & $100.0$ & $7.863$ & $150.0$ & $%
11.641$ \\
$n=2$ &  & $0.324$ &  & $0.380$ &  & $0.533$ &  & $0.749$ \\
$n=3$ &  & $0.138$ &  & $0.131$ &  & $0.142$ &  & $0.172$%
\end{tabular}%
\end{table}


\newpage

\begin{thebibliography}{999}
\bibitem{1} H. P. Nills, Phys. Rep. 110, 1 (1984).

\bibitem{2} H. Haber and G. Kane, Phys. Rep. 117, 75 (1985).

\bibitem{3} ALEPH Collab. (R. Barate \textit{et al.}), Phys. Lett. B 434,
437 (1998).

\bibitem{4} OPAL Collab. (G. Abbiendi \textit{et al.)}, Eur. Phys. J. C 20,
601 (2001).

\bibitem{5} OPAL Collab. (G. Abbiendi \textit{et al.)},Phys. Lett. B 456, 95
(1999).

\bibitem{6} L3 Collab. (M. Acciarri \textit{et al.}), Phys. Lett. B 471, 308
(1999).

\bibitem{7} DELPHI Collab. (P. Abreu \textit{et al.}), Eur. Phys. J. C 6,
385 (1999).

\bibitem{8} DELPHI Collab. (P. Abreu \textit{et al.}), Phys. Lett. B 444,
491 (1998).

\bibitem{9} M. Drees and K. Hikasa, Phys. Lett. B 252, 127 (1990).

\bibitem{10} A. Barti \textit{et al.}, Z. Phys. C 73, 469 (1997).

\bibitem{11} DO Collab. (S. Abachi \textit{et al}.), Phys. Rev. Lett. 76,
2222 (1996).

\bibitem{12} CDF Collab. (T. Affolder \textit{et al}.), Phys. Rev. Lett. 84,
5273 (2000).

\bibitem{13} P. R. Harrison and C. H. L. Smith, Nucl. Phys. B 213, 223
(1983).

\bibitem{14} E. Reya and D. P. Roy, Phys. Rev. D 32, 645 (1985).

\bibitem{15} M. Carena \textit{et al.}, Phys. Rev. Lett. 86, 4463 (2001).

\bibitem{16} E. L. Berger \textit{et al.}, Phys. Rev. Lett. 86, 4231 (2001)
[arXiv: hep-ph/0012001].

\bibitem{17} J. Cao, Z. Xiong and J. M. Yang, Phys. Rev. Lett. 88, 111802
(2002).

\bibitem{18} G. -C. Cho, Phys. Rev. Lett. 89, 091801 (2002) [arXiv:
hep-ph/0204348].

\bibitem{19} S. -W. Baek, Phys. Lett. B 541, 161 (2002) [arXiv:
hep-ph/0205013].

\bibitem{20} CLEO Collab. (V. Savinov \textit{et al}.), Phys. Rev. D 63,
051101 (2001).

\bibitem{21} G. Talor, EPC presentation, 20/07/2000, www.cern.ch.

\bibitem{22} S. Pacetti and Y. Srivastava, arXiv: hep-ph/0007318.

\bibitem{23} H. Baer, K. Cheung and J. Gunion, Phys. Rev. D 59, 075002
(1999).

\bibitem{24} Kingman Choung and Wai-Yee Keung, Phys. Rev. Lett. 89, 221801
(2002) [arXiv: hep-ph/0205345].

\bibitem{25} Edmond L. Berger and Jungil Lee, Phys. Rev. D 65, 114003 (2002)
[arXiv: hep-ph/0203092].

\bibitem{26} Adam K. Leibovich and David L. Rainwater, Phys. Rev. Lett. 88,
221801 (2002) [arXiv: hep-ph/0202174].

\bibitem{27} E. Berger and L. Clavelli, Phys. Lett. B 512, 115 (2001).

\bibitem{28} T. Becher \textit{et al.}, Phys. Lett. B 540, 278 (2002)
[arXiv: hep-ph/0205274].

\bibitem{29} U. Nierste and T. Plehn, Phys. Lett. B 493, 104 (2000).

\bibitem{30} E. L. Berger \textit{et al.}, Phys. Rev. D 66, 095001 (2002)
[arXiv: hep-ph/0205342].

\bibitem{31} T. Becher \textit{et al.}, Budapest 2001, High Energy Physics,
hep2001/090 [arXiv: hep-ph/0112129].

\bibitem{32} A. Dedes and H. K. Dreiner, JHEP 0106, 006 (2001) [arXiv:
hep-ph/0009001].

\bibitem{33} S. M. Ikhdair and R. Sever, Int. J. Mod. Phys. A 21, 3989
(2006).

\bibitem{34} L. Motyka and K. Zalewiski, Eur. Phys. J. C 4,\textbf{\ }107
(1998).

\bibitem{35} L. Motyka and K. Zalewiski, Z. Phys. C 69,\textbf{\ }343 (1996).

\bibitem{36} S. J. Gates, Jr. and Oleg Lebedev, Phys. Lett. B 477, 216
(2000) [arXiv: hep-ph/9912362].

\bibitem{37} R. Peschanski, Nucl. Phys. Proc. Suppl. 86, 170 (2000) [arXiv:
hep-ph/9909359].

\bibitem{38} S. M. Ikhdair and R. Sever\textit{,} Z. Phys. C 56, 155 (1992).

\bibitem{39} S. M. Ikhdair and R. Sever\textit{,} Z. Phys. C 58, 153 (1993).

\bibitem{40} S. M. Ikhdair and R. Sever\textit{,} Z. Phys. D 28, 1 (1993).

\bibitem{41} S. M. Ikhdair and R. Sever\textit{, }Int. J. Mod. Phys. A 18,
4215 (2003).

\bibitem{42} S. M. Ikhdair and R. Sever\textit{, }Int. J. Mod. Phys. A 20,
4035 (2005).

\bibitem{43} S. M. Ikhdair and R. Sever\textit{, }Int. J. Mod. Phys. A 21,
2191 (2006).

\bibitem{44} S. M. Ikhdair and R. Sever\textit{, }Int. J. Mod. Phys. A 21,
6699 (2006).

\bibitem{45} S. Ikhdair \textit{et al.,} Tr. J. Phys. 16, 510 (1992).

\bibitem{46} S. Ikhdair \textit{et al.,} Tr. J. Phys. 17, 474 (1993).

\bibitem{47} S. M. Ikhdair and R. Sever\textit{, }Int. J. Mod. Phys. A 19,
1771 (2004).

\bibitem{48} S. M. Ikhdair and R. Sever\textit{, }Int. J. Mod. Phys. A 20,
6509 (2005).

\bibitem{49} S. M. Ikhdair and R. Sever\textit{, }Int. J. Mod. Phys. E 17,
669 (2008).

\bibitem{50} E. Eichten \textit{et al}., Phys. Rev. D 17, 3090 (1978).

\bibitem{51} E. Eichten \textit{et al}., Phys. Rev.D 21, 203 (1980).

\bibitem{52} R. L. Jaffe and K. Johnson, Comments Nucl. Part. Phys. 7, 107
(1977).

\bibitem{53} Y. Nambu, Phys. Rev. D 10, 4262 (1974).

\bibitem{54} K. D. Born \textit{et al}., Phys. Lett. B 329, 325 (1994).

\bibitem{55} Particle Data Group Collab. (W. M. Yao \textit{et al}.), J.
Phys. G 33, 1 (2006).

\bibitem{56} T. Imbo \textit{et al.,} Phys. Rev. D\textbf{\ }29, 1669 (1984).

\bibitem{57} H. Christiansen \textit{et al.,} Phys. Rev. A 40, 1760 (1989).

\bibitem{58} T. D. Imbo and U. P. Sukhatme, Phys. Rev. Lett. 54, 2184 (1985).

\bibitem{59} T. Barnes, S. Godfrey and E. S. Swanson, Phys. Rev. D 72,
054026 (2005).

\bibitem{60} T. Barnes and G. I. Ghandour, Phys. Lett. B 118, 411 (1982).

\bibitem{61} Yu-Qi Chen and Yu-Ping Kuang, Phys. Rev. D 46, 1165 (1992).

\bibitem{62} Yu-Qi Chen, Yu-Ping Kuang and Robert J. Oakes, Phys. Rev. D 52,
264 (1995).

\bibitem{63} Yu-Qi Chen, and Robert J. Oakes, Phys. Rev. D 53, 5051 (1996).

\bibitem{64} A. Billoire \textit{et al}., Nucl. Phys. B (Proc. Suppl.) 4,
199 (1988).

\bibitem{65} A. F. Falk, B. Grinstein and M. E. Luke, Nucl. Phys. B 357, 185
(1991).

\bibitem{66} E. Eichten and F. Feinberg, Phys. Rev. Lett. 43, 1205 (1979).

\bibitem{67} E. Eichten and F. Feinberg, Phys. Rev. D 23, 2724 (1982).

\bibitem{68} J. Pantaleone, S. -H. H. Tye and Y. J. Ng, Phys. Rev. D 33, 777
(1986).

\bibitem{69} S. N. Gupta and S. F. Radford, Phys. Rev. D 24, 2309 (1981).

\bibitem{70} S. N. Gupta, S. F. Radford and W. W. Repko, Phys. Rev. D 26,
3305 (1982).

\bibitem{71} E. J. Eichten and C. Quigg, Phys. Rev. D 49, 5845 (1994).

\bibitem{72} L. P. Fulcher, Z. Chen, and K. C. Yeong, Phys. Rev. D 47, 4122
(1993).

\bibitem{73} L. P. Fulcher, Z. Chen, and K. C. Yeong, Phys. Rev. D 50, 447
(1994).

\bibitem{74} L. P. Fulcher, Phys. Rev. D 60, 074006 (1999).

\bibitem{75} L. P. Fulcher, Phys. Rev. D 44, 2079 (1991).

\bibitem{76} H. J. Schnitzer, Phys. Rev. D 13, 74 (1975).

\bibitem{77} M. R. Arafah \textit{et al.}, Annals of Phys. 220, 55 (1992).

\bibitem{78} W. Lucha \textit{et al.}, Phys. Rev. D 46, 1088 (1992).

\bibitem{79} W. Lucha \textit{et al.}, D 45, 385 (1992).

\bibitem{80} W. Lucha \textit{et al.}, D 45, 1233 (1992).

\bibitem{81} W. Lucha, F. F. Sch\"{o}berl, and D. Gromes, Phys. Rep. 200,
127 (1991).

\bibitem{82} C. Quigg and J. L. Rosner, Phys. Lett. B 71,153 (1977).

\bibitem{83} C. Quigg and J. L. Rosner, Phys. Rep. 56, 167 (1979).

\bibitem{84} L. I. Abou-Salem, Int. J. Mod. Phys. A 20, 4113 (2005).

\bibitem{85} M. Baldicchi and G. M. Prosperi, Phys. Lett. B 436, 145 (1998).

\bibitem{86} M. Baldicchi and G. M. Prosperi, Phys. Rev. D 62, 114024 (2000).

\bibitem{87} M. Baldicchi and G. M. Prosperi, Fiz. B 8, 251 (1999).

\bibitem{88} C. R. Nappi, Phys. Rev. D 25, 84 (1982).

\bibitem{89} J. A. Grifols and A. M\'{e}ndez, Phys. Lett. B 144, 123 (1984).

\bibitem{90} E. Eichten \textit{et al.}, Rev. Mod. Phys. 80, 1161 (2008)
[arXiv: hep-ph/0701208].

\bibitem{91} Belle Collab. (S. K. Choi \textit{et al}.), Phys. Rev. Lett.
89, 102001 (2002).

\bibitem{92} Belle Collab. (S. K. Choi \textit{et al}.), Phys. Rev. Lett.
89, 129901 (2002) [arXiv: hep-ex/0206002].

\bibitem{93} BABAR Collab. (B. Aubert \textit{et al}.), Phys. Rev. Lett. 92,
142002 (2004) [arXiv: hep-ex/0311038].

\bibitem{94} CLEO Collab. (D. M. Asner \textit{et al}.), Phys. Rev. Lett.
92, 142001 (2004) [arXiv: hep-ex/0312058].

\bibitem{95} Belle Collab. (S. K. Choi \textit{et al}.), Phys. Rev. Lett.
91, 262001 (2003) [arXiv: hep-ex/0309032].

\bibitem{96} CDF II Collab. (D. Acosta \textit{et al}.), Phys. Rev. Lett.
93, 072001 (2004) [arXiv: hep-ex/0312021].

\bibitem{97} N. A. Tornqvist, arXiv: hep-ph/0308277.

\bibitem{98} F. E. Close and P. R. Page, Phys. Lett. B 578, 119 (2004)
[arXiv: hep-ph/0309253].

\bibitem{99} E. S. Swanson, Phys. Lett. B 588, 189 (2004) [arXiv:
hep-ph/0311229].

\bibitem{100} T. Barnes and S. Godfrey, Phys. Rev. D 69, 054008 (2004)
[arXiv: hep-ph/0311162].

\bibitem{101} E. J. Eichten, K. Lane and C. Quig, Phys. Rev. D 69, 094019
(2004) [arXiv: hep-ph/0401210].

\bibitem{102} K. Abe, arXiv: hep-ex/0505037.

\bibitem{103} K. Abe, arXiv: hep-ex/0505038.

\bibitem{104} Kamal K. Seth, arXiv: hep-ex/0501022.

\bibitem{105} A. Tomaradze, J. Phys. Conf. Ser. 9,119 (2005) [arXiv:
hep-ex/0410090].

\bibitem{106} Jonathan L. Rosner, Annals Phys. 319, 1 (2005)
[arXiv:hep-ph/0411003].

\bibitem{107} S. N. Gupta and J. M. Johnson, Phys. Rev. D 54, 2075 (1996).

\bibitem{108} S. Narison, Phys. Lett. B 387, 162 (1996).

\bibitem{109} M. Chabab, Phys. Lett. B 325, 205\textbf{\ }(1994).

\bibitem{110} C. T. H. Davies \textit{et al.}, Phys. Lett. B 382, 131 (1996).

\bibitem{111} CP-PACS Collaboration (M. Okamoto \textit{et al}.), Phys. Rev.
D 65, 094508 (2002).

\bibitem{112} X. Liao and T. Manke, Phys. Rev. D 65, 074508 (2002).

\bibitem{113} S. Recksiegel and Y. Sumino, Phys. Lett. B 578, 369 (2004).

\bibitem{114} S. Recksiegel and Y. Sumino, Phys. Rev. D 67, 014004 (2003).

\bibitem{115} W. Kwong and J. Rosner, Phys. Rev. D 44, 212 (1991).

\bibitem{116} W. Kwong and J. Rosner, Phys. Rev. D 38, 279 (1988).

\bibitem{117} P. Colangelo and F. De Fazio, Phys. Rev. D 61, 034012 (2000).

\bibitem{118} M. Baker, J.S. Ball, and F. Zachariasen, Phys. Rev. D 45, 910
(1992).

\bibitem{119} R. Roncaglia \textit{et al.,} Phys. Rev. D 32, 189 (1985).

\bibitem{120} S. Godfrey, Phys. Rev. D 70, 054017 (2004).

\bibitem{121} E. Bagan \textit{et al.}, CERN Report No. TH. 7141/94
(unpublished).

\bibitem{122} N. Brambilla and A. Vairo, Phys. Rev. D 62, 094019 (2000).

\bibitem{123} S. Gershtein \textit{et al}., Phys. Rev. D 51, 3613 (1995).

\bibitem{124} D. Ebert, R. N. Faustov and V. O. Galkin, Phys. Rev. D 67,
014027 (2003).

\bibitem{125} D. Ebert \textit{et al}.\textit{, }Mod. Phys. Lett. A 17, 803
(2002).

\bibitem{126} D. Ebert \textit{et al}.\textit{,} Phys. Rev. D 67, 014027
(2003) [arXiv:hep-ph/0210381].

\bibitem{127} D. Ebert \textit{et al}.\textit{,} AIP Conf. Proc. 619, 336
(2002) [arXiv:hep-ph/0110190].

\bibitem{128} L. Haysak \textit{et al}.,\ Czech. J. Phys. 55, 541 (2005)
[arXiv: hep-ph/0301188].

\bibitem{129} X. Liao and T. Manke, Phys. Rev. D 65, 074508 (2002).

\bibitem{130} C. -H. Chang, J. Y. Cui and J. M. Yang, Commun. Theor. Phys.
39, 197 (2003) [arXiv:hep-ph/0211164].
\end{thebibliography}
\end{document}